\documentclass[fleqn,usenatbib]{mnras}

\newcommand{\dmunits}{\ensuremath{{\rm pc \, cm^{-3}}}}
\newcommand{\dmcosmic}{\ensuremath{{\rm DM}_{\rm cosmic}}}

\newcommand{\dmhost}{\ensuremath{{\rm DM}_{\rm host}}}

\newcommand{\dmhalo}{\ensuremath{{\rm DM}_{\rm halo}}}
\newcommand{\dmism}{\ensuremath{{\rm DM}_{\rm ISM}}}

\newcommand{\dmne}{\ensuremath{{\rm DM}_{\rm NE2001}}}
\newcommand{\dmymw}{\ensuremath{{\rm DM}_{\rm YMW16}}}
\usepackage{newtxtext,newtxmath}

\usepackage[T1]{fontenc}
\usepackage{comment}

\DeclareRobustCommand{\VAN}[3]{#2}
\let\VANthebibliography\thebibliography
\def\thebibliography{\DeclareRobustCommand{\VAN}[3]{##3}\VANthebibliography}

\usepackage{graphicx}	
\usepackage{amsmath}	
\usepackage{xcolor}

\usepackage{csvsimple} 

\newcommand{\sched}{\texttt{Sched}}
\newcommand{\difx}{\texttt{DiFX}}
\newcommand{\casa}{\texttt{CASA}}
\newcommand{\xgpu}{\texttt{xGPU}}
\newcommand{\wsclean}{\texttt{WSClean}}
\newcommand{\pybdsf}{\texttt{PyBDSF}}

\title[Transient buffer FRB localisation]{A study of two FRBs with low polarization fractions localized with the MeerTRAP transient buffer system}

\author[K. M. Rajwade et al.]{K.~M.~Rajwade,$^{1,2}$\thanks{E-mail: kaustubh.rajwade@physics.ox.ac.uk}
L.~N.~Driessen,$^{3}$\thanks{E-mail:laura.driessen@sydney.edu.au}\thanks{both authors contributed equally}
E.~D.~Barr,$^{4}$
I.~Pastor-Marazuela,$^{5}$
M.~Berezina,$^{4,16}$
F.~Jankowski,$^{7,5}$
\newauthor
A.~Muller,$^{6}$
L.~Kahinga,$^{8}$
B.~W.~Stappers,$^{5}$
M.~C. Bezuidenhout,$^{9,10}$
M.~Caleb,$^{3}$
A. Deller,$^{11}$
W.~Fong,$^{12}$
\newauthor
A.~Gordon,$^{12}$
M.~Kramer,$^{4}$
M.~Malenta,$^{5}$
V.~Morello,$^{13}$
J. X.~Prochaska,$^{8}$
S.~Sanidas,$^{5}$
M.~Surnis,$^{14}$
N.~Tejos$^{15}$
\newauthor
S. Wagner$^{16}$
\\
$^{1}$ Astrophysics, University of Oxford, Denys Wilkinson Building, Keble Road , Oxford OX1 3RH, UK \\
$^{2}$ ASTRON, 4 Oude Hoogeveensedijk, Dwingeloo 7991 PD, The Netherlands\\
$^{3}$ Sydney Institute for Astrophysics, School of Physics, University of Sydney, NSW, Sydney 2006, Australia\\
$^{4}$ Max-Planck-Institut f\"{u}r Radioastronomie, Auf dem H\"{u}gel 69, D-53121 Bonn, Germany\\
$^{5}$ Jodrell Bank Centre for Astrophysics, University of Manchester, Oxford Road, Manchester M13 9PL\\
$^{6}$ Maria Mitchell Observatory, Nantucket, MA 02554, USA\\
$^{7}$ LPC2E, Universit\'{e} d'Orl\'{e}ans, CNRS, 3A Avenue de la Recherche Scientifique, 45071 Orl\'{e}ans, France\\
$^{8}$ University of California, Santa Cruz, 1156 High St., Santa Cruz, CA 95064, USA\\
$^{9}$ Department of Mathematical Sciences, University of South Africa, Cnr Christiaan de Wet Rd and Pioneer Avenue, Florida Park, 1709, Roodepoort, South Africa\\
$^{10}$ Centre for Space Research, North-West University, Potchefstroom 2531, South Africa\\
$^{11}$ Centre for Astrophysics and Supercomputing, Swinburne University of Technology, Hawthorn, VIC 3122, Australia\\
$^{12}$ Center for Interdisciplinary Exploration and Research in Astrophysics (CIERA), Department of Physics and Astronomy,\\
Northwestern University, Evanston, IL 60208, USA\\
$^{13}$ SKA Observatory, Jodrell Bank, Lower Withington, Macclesfield, SK119FT, UK\\
$^{14}$ Department of Physics, IISER Bhopal, Bhauri Bypass Road, Bhopal, 462066, India\\
$^{15}$ Instituto de F\'isica, Pontificia Universidad Cat\'olica de Valpara\'iso, Casilla 4059, Valpara\'iso, Chile\\
$^{16}$ Landessternwarte, Universit\"{a}t Heidelberg, K\"{o}nigstuhl 12, D-69117 Heidelberg, Germany \\
}

\date{Accepted XXX. Received YYY; in original form ZZZ}

\pubyear{2024}

\begin{document}
\label{firstpage}
\pagerange{\pageref{firstpage}--\pageref{lastpage}}
\maketitle

\begin{abstract}
Localisation of fast radio bursts (FRBs) to arcsecond and sub-arcsecond precision maximizes their potential as cosmological probes. To that end, FRB detection instruments are deploying triggered complex-voltage capture systems to localize FRBs, identify their host galaxy and measure a redshift. Here, we report the discovery and localisation of two FRBs (20220717A and 20220905A) that were captured by the transient buffer system deployed by the MeerTRAP instrument at the MeerKAT telescope in South Africa. We were able to localize the FRBs to precision of $\sim$1 arc-second that allowed us to unambiguously identify the host galaxy for FRB~20220717A (posterior probability$\sim$0.97). FRB~20220905A lies in a crowded region of the sky with a tentative identification of a host galaxy but the faintness and the difficulty in obtaining an optical spectrum preclude a conclusive association. The bursts show low linear polarization fractions (10--17$\%$) that conform to the large diversity in the polarization fraction observed in apparently non-repeating FRBs akin to single pulses from neutron stars. We also show that the host galaxy of FRB~20220717A contributes roughly 15$\%$ of the total dispersion measure (DM), indicating that it is located in a plasma-rich part of the host galaxy which can explain the large rotation measure. The scattering in FRB~20220717A can be mostly attributed to the host galaxy and the intervening medium and is consistent with what is seen in the wider FRB population.
\end{abstract}

\begin{keywords}
radio continuum: transients -- stars: neutron -- techniques: interferometric
\end{keywords}



\section{Introduction}
Fast radio bursts (FRBs) are intense, millisecond-duration radio flashes that originate from cosmological distances~\citep{lorimer2007L}. They have remained one of the most enigmatic astrophysical mysteries since their discovery over a decade ago. Several theories have been proposed to explain their origin but we still lack any definitive evidence to decipher their nature. The detection of repeating FRBs allowed astronomers to regularly monitor the sources and enable precise localisation to their host galaxies~\citep{tendulkar2017}. These follow-up studies have been important to put constraints on their progenitors. The discovery of FRB-like bursts from a Galactic magnetar SGR J1935+2154 suggests that highly magnetized neutron stars (magnetars) have the ability to produce luminous radio bursts~\citep{SGR1935chime,bochenek2020}. This suggested that we should expect FRBs in star forming regions of their host galaxies where most of the magnetars are produced via core-collapse supernovae. This conjecture was put to the test again when a repeating FRB was discovered and localised to a globular cluster in a near-by galaxy M81~\citep{kirsten2022}. One needs to invoke exotic models for the creation of magnetars in an environment that is dominated by an old stellar population. These results already show the importance of precise localisations of FRBs and their environs that provide important clues about their progenitors. Moreover, it also could help in determining the distribution of FRBs across different galaxy types, probe the intergalactic medium with extreme precision and count the `missing' baryons and their distribution~\citep{macquart2020}. All of these advancements can lead to a deeper understanding of the physics behind these enigmatic signals.

Until a few years ago, precise localisation of the FRBs was only possible with repeating FRBs as it allows for regular follow-up using radio interferometers. However, recent advancements in instrumentation and observing strategies have enabled arc-second localisations of one-off FRBs, opening up the field entirely. The most significant breakthrough in localising single FRBs came with the development of the commensal real-time ASKAP FAST Transients survey~\citep[CRAFT;][]{bannister2018}. CRAFT enabled ASKAP to detect and localise FRBs in real-time, providing rapid follow-up optical observations  and identification of host galaxies. Since then, other radio telescopes have followed suit and are now spear-heading the real-time localisation efforts of one-off FRBs~\citep{chimefrbpipeline, bannister2018, ravi2023}. In this paper, we report two sub-arcsecond localisations of FRBs using the transient buffer capture mode on MeerTRAP: A commensal, real-time FRB detector at the MeerKAT telescope in South Africa. The paper is organized as follows; in section 2, we describe the transient buffer capture system. In section 3, we describe the discovery, localisation and optical follow-up of the first two FRBs with this system. In section 4, we discuss the properties of the FRBs and their host galaxies and in section 5, we summarize our results and conclusions.

\section{MeerTRAP Transient Buffer System}

\begin{figure*}
	\includegraphics[width=\textwidth]{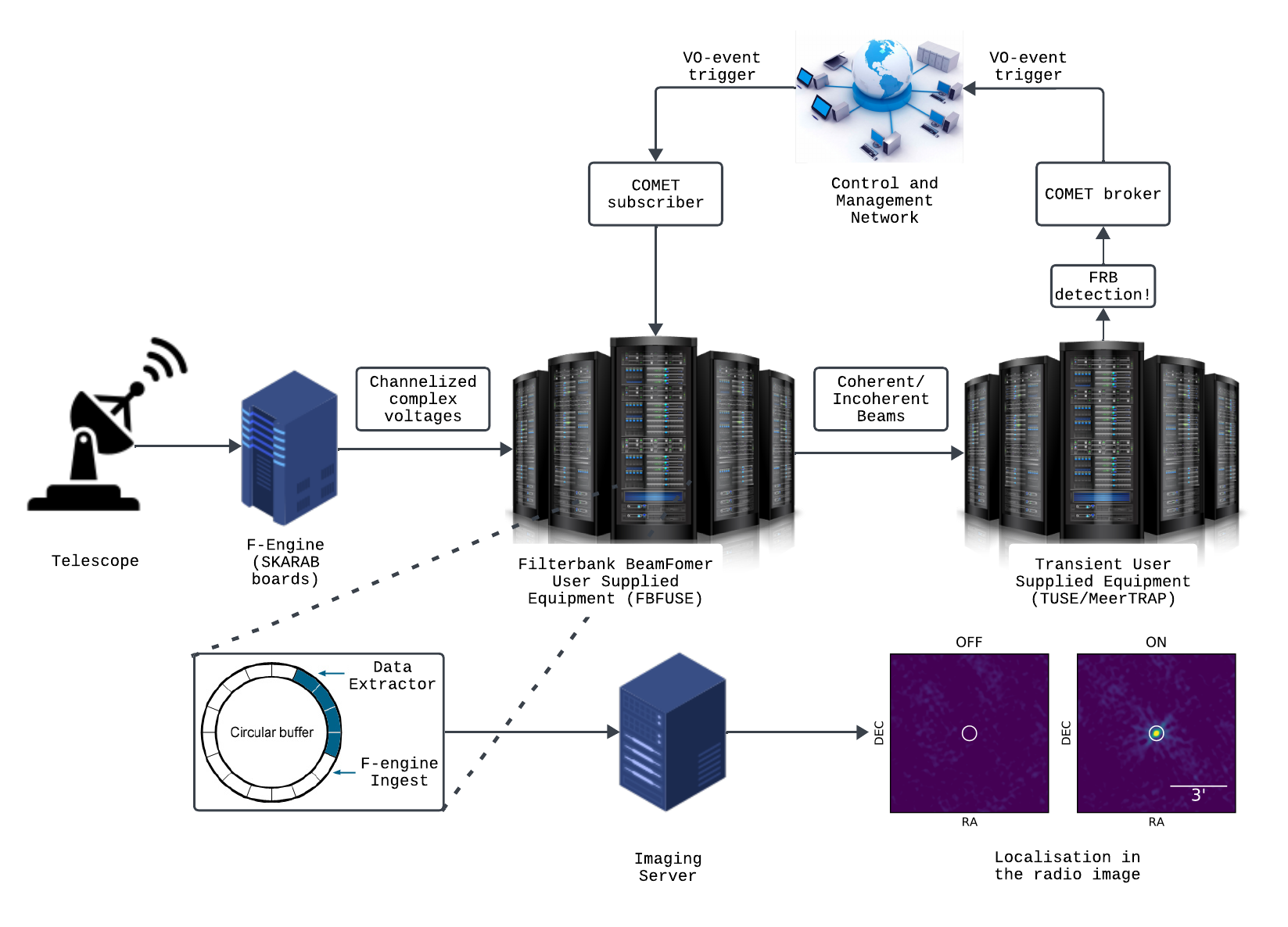}
    \caption{A flow chart showing the entire transient buffer trigger pipeline (see text for details). Here the F-engine corresponds to where the poly-phase filter is applied on the complex voltages streaming from the telescope. 
    The figure has been created using \textsc{Lucid Chart}.}
    \label{fig:flow_diagram}
\end{figure*}

\subsection{The real-time search}
The detailed description of the real-time FRB detection system has been presented in~\cite{rajwade2021} and~\cite{rajwade2022}. Figure~\ref{fig:flow_diagram} shows the detailed flow diagram of the system. Raw data from each antenna are channelized using a poly-phase filter~\citep{byl2020} to create a discretely channelized complex voltage datastream. This datastream is acquired by the Filterbank BeamFomer User Supplied Equipment (FBFUSE) where these data are detected and converted into total power beams across the FoV of MeerKAT. MeerTRAP observations typically use only the inner 40 dishes of the MeerKAT array for beamforming. This is a trade-off between sensitivity and achievable field of view (FoV) given the finite compute resources available to FBFUSE \citep{chen2021}. Even when only beamforming a subset of the antennas, FBFUSE ingests the full complement of channelised voltages from the MeerKAT antennas. This is essential for the operation of the transient buffer. The Transient User Supplied Equipment (TUSE) receives the coherent total power beams from FBFUSE and runs a real-time search on the data for FRBs and other transients. 

\subsection{Detection and trigger}

In order to save complex voltages from the telescope, it is important to send out prompt triggers to the beamformer immediately after the detection of an FRB to initiate data extraction. Typically, the real-time system has to process the data, classify the candidates and send a trigger within 45~seconds of receiving the data from the beamformer. To that end, we decided to use low-latency VOEvent alerts to communicate triggers. That is because VOEvents are well established in the transient community, a software ecosystem exists, a VOEvent standard for FRB alerts had already been proposed \citep{petroff2017}, and was subsequently adopted at several radio telescopes, most notably CHIME. For MeerTRAP, we implemented a VOEvent-based software to trigger the voltage buffer read-out on the FBFUSE cluster from the real-time transient detection system running on the TUSE servers \citep{jankowski2022vo}. VOEvent messages are in XML format \citep{seaman2011} and contain the parameters of the alert, e.g.\ a unique identifier, the author, the event time, its sky position, and the instrumental setup. The event packets are distributed by brokers, for which we employ the \textsc{comet} software \citep{swinbank2014}, both locally on the MeerTRAP head nodes and the central MeerKAT observatory-wide broker. A containerised \textsc{comet} subscriber runs on the FBFUSE head node, waiting for events. When an FRB, or any other transient, is detected by the MeerTRAP pipeline, its parameters are written into a VOEvent message which is sent to the local \textsc{comet} broker and forwarded to the observatory-wide one. The alert is then received by the FBFUSE subscriber which parses the contents and converts them into a request to write-out the corresponding complex voltage data from the transient buffer. More details are presented in \citet{jankowski2022vo} and software are available online\footnote{\url{https://github.com/fjankowsk/meertrig/}}. Using VOEvents has the advantage that we can easily disseminate our triggers to external collaborators in the future.

\subsection{Extraction of complex voltage data and phase-up}
\label{sec:extraction}
Data from MeerKAT channelisers arrive on the the FBFUSE cluster as a 1.8 Tb/s Ethernet stream, split over 256 multicast groups, with each group containing 1/256$^{\rm th}$ of the full MeerKAT bandwidth for all the available antennas included in the current observations. The groups are split such that each of the 32-servers that comprise the FBFUSE cluster ingests 8 groups, 4 per network interface. Physically, the processing for each set of 4 multicast groups is mapped to a single non-uniform memory architecture (NUMA) node, hosting a network card, CPU, GPU and 192 GB of DDR4 RAM. The depth of the transient buffer that can be accommodated on such a system is determined by $t_{\rm tb} = 8M / (2 N_{\rm pol} N_{\rm ant} B N_{\rm bits})\, \mathrm{s}$, where $M$ is the available memory in bytes, $N_{\rm pol}$ is the number of polarisations, $N_{\rm ant}$ is the number of antennas being ingested, $B$ is the received bandwidth per NUMA node in Hz and $N_{\rm b}$ is the bit depth per sample. For MeerKAT, we have $N_{\rm pol} = 2$, $N_{\rm bits} = 8$, $N_{\rm ant} \leq 64$ and $B = 8.5$, $13.375$, or $13.671875$ MHz at UHF (816~MHz), L-band (1.4~GHz), and S-band (2.2~GHz), respectively. Approximately 95~\% of the RAM ($\sim$182~GB) on each FBFUSE NUMA node is available for the transient buffer, hence we achieve a buffer depth of $\sim$54, 56 and 88~s at L-band (1284 MHz), UHF (816~MHz) and S-band (2500~MHz) for the full array. The buffer depth may be increased by moving to a lower bandwidth receiver or by specifying that only a subset of the current sub-array be used (although it should be noted that the subset used by the transient buffer defines the superset available for beamforming). Depending on the number of frequency channels requested from the MeerKAT correlator, the time resolution of the transient buffer data varies from 1.9 to 36~$\mu$s.

\begin{figure*}
        \centering
	\includegraphics[width=0.40\textwidth]{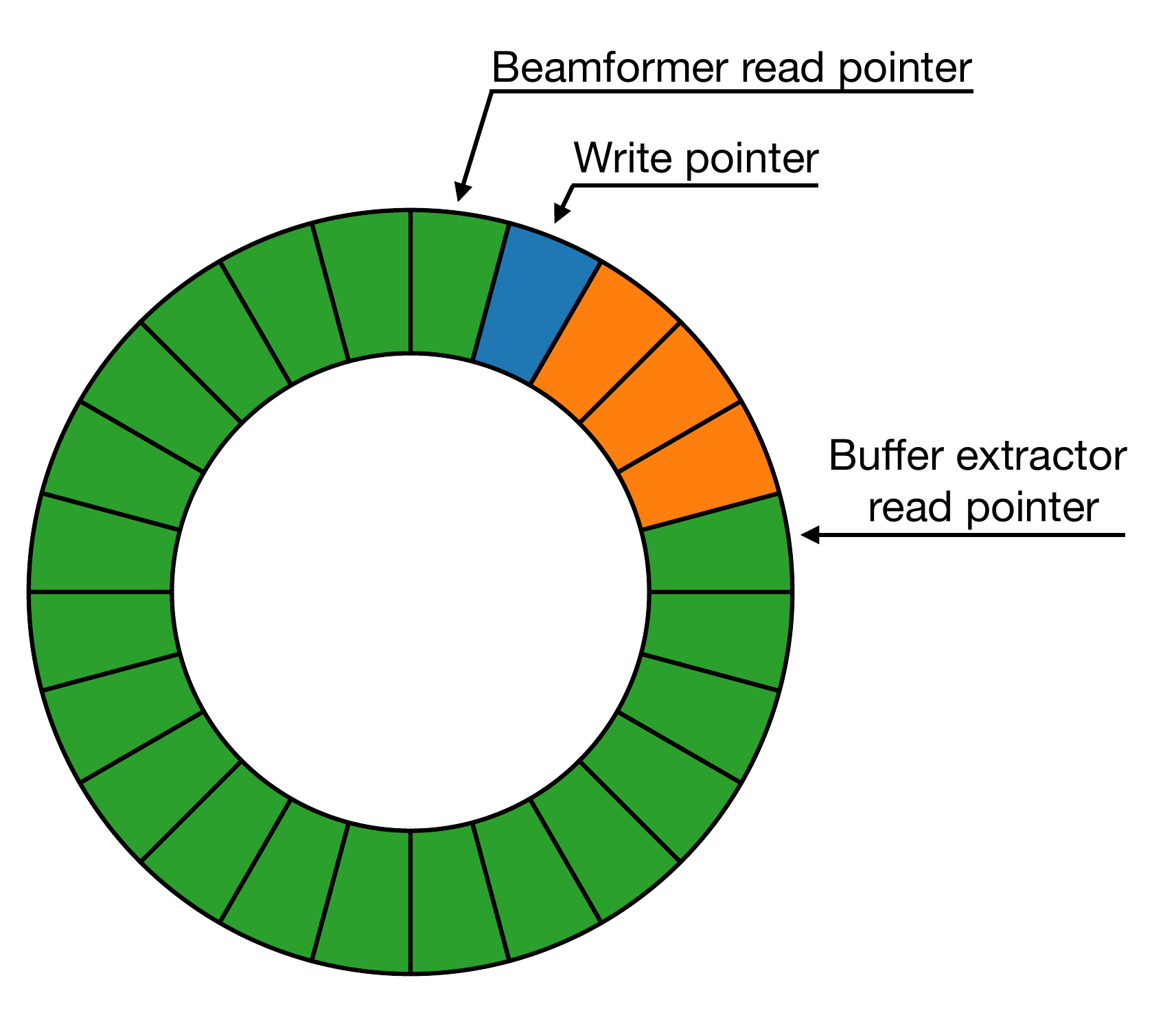}
\includegraphics[width=0.54\textwidth]{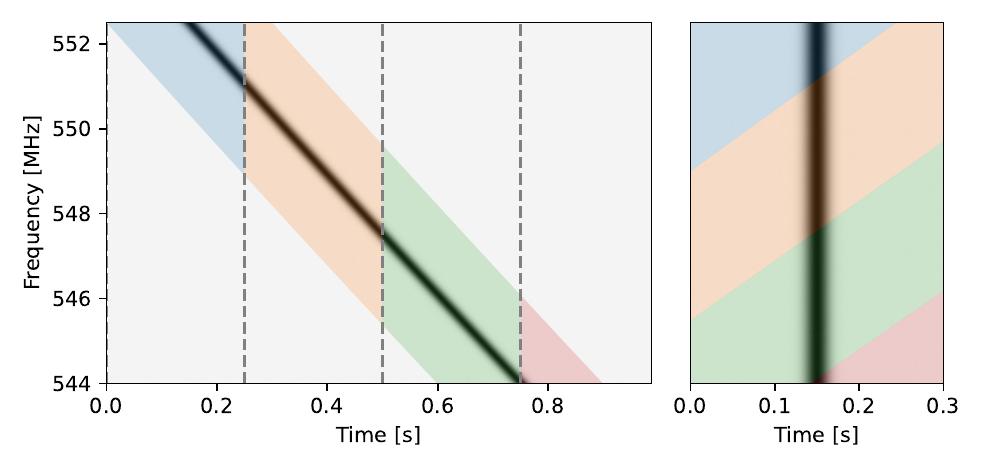}
        \caption{\textbf{Left:} Shared memory ring buffer configuration for the FBFUSE transient buffer. Each segment represents a block of memory in the ring buffer, with blue showing a block that is being written to, green showing blocks that are occupied and orange showing blocks that free and can be written to. Shown are the positions of the write pointer for data coming from the MeerKAT correlator network, the beamformer read pointer for data going through the FBFUSE beamforming pipeline and the buffer extractor read pointer for data being recorded upon receipt of a trigger. The write pointer progresses though the buffer in a clock-wise direction. \textbf{Right:} The algorithm that extracts the data corresponding to the DM of the detected FRB after accounting for the dispersion delay. The dispersion delay has been shown here as a linear trend for simplicity. The FRB data are spread accross several data blocks due to the delay as shown by the dashed vertical lines. The coloured regions shows the data that are extracted from each data block.}
    \label{fig:smrb}
\end{figure*}

As illustrated in Figure \ref{fig:smrb}, the FBFUSE transient buffer is implemented as a \textsc{PSRDADA}\footnote{\url{https://psrdada.sourceforge.net/}} shared memory ring buffer (SMRB) with one writer and two readers. The writing process captures data from the MeerKAT correlator network, orders it by time, antenna and frequency and writes it to the SMRB. The primary reading process is the beamformer itself, which operates in real time, consuming and processing blocks from the SMRB as they become available. The secondary reading process is the transient buffer data extractor. This process does not immediately read blocks from the SMRB but instead monitors the overall usage of the SMRB and holds open blocks in the buffer, only releasing them when the overall occupancy of the buffer reaches 95\%. It thus guarantees that at least 95\% of the buffer is maintained in memory at all times. The remaining 5\% of the buffer is required to be left unoccupied to allow sufficient time for data extraction and processing on receipt of a trigger event (see below) such that the writing process is not blocked, resulting in data loss.

The triggers received by FBFUSE are propagated to the buffer data extractor process via a UNIX socket. Each is formatted as a JSON message containing a DM, reference frequency, start UTC, end UTC and trigger identifier. The start and end UTC along with the reference frequency and DM define the section of data to be extracted from the transient buffer. As noted above, extraction of data from the SMRB must be sufficiently fast as to avoid blocking the writing process. Several tests have shown that the instrument can safely write up to 300 ms of the buffer to disk at a time without affecting the capture of data from the MeerKAT correlator network. 
As 300 ms may be shorter than duration of the time delay of a highly dispersed FRB, the buffer data extractor incoherently dedisperses the buffer data at the time of extraction. Upon receipt of a trigger, the buffer data extractor re-references the start and end UTCs of the trigger to the highest frequency in the currently processed subband and scans through the buffer until it reaches the block containing the start of the event. The frequency channels and times corresponding to the event window are then extracted for all antennas and polarisations and written to a temporary memory buffer in dedispersed order. This process continues over subsequent blocks until the end of the event is reached, at which point the temporary memory buffer is written to disk with a header containing observation and trigger metadata. This process is illustrated in the right-hand panel of Figure \ref{fig:smrb}.

In order to aide in the downstream analysis of the extracted voltages, FBFUSE records a snapshot of the current complex gain solutions as calculated by the MeerKAT Science Data Processor~\citep{jonas2016}. These are written locally as \textsc{Numpy} arrays to be applied to the transient buffer data extracted for any FRB.

\subsection{Imaging and localisation}
\subsubsection{Producing measurement sets}

The extracted transient buffer data are correlated using \texttt{xGPU}\footnote{\texttt{xGPU}: \url{https://github.com/GPU-correlators/xGPU}} \citep{clark_accelerating_2011}. These data already have the geometric delays applied and we apply the gain and phase solutions to each antenna, time and frequency channel to phase-up the data to the pointing centre of the observation using the solutions obtained during the initial delay calibration. Each file produced contains one subband (1/64$^{\rm th}$ of the full bandwidth) and, due to dispersion correction, has a different start time. In order to calibrate and image the correlated visibilities, they need to be packaged with appropriate metadata (e.g., phase centre position, baseline direction cosines etc.) in a recognised visibility file format such as FITS-Interferometry Data Interchange (IDI) format~\footnote{\url{https://lweb.cfa.harvard.edu/~jzhao/SMA-FITS-CASA/docs/AIPSMEMO102.pdf}} or a Measurement Set (MS). We made use of the \textsc{difx2fits} application provided by \difx~\citep{deller_difx_2007, deller_difx-2_2011} to produce FITS-IDI files that could subsequently be converted to an MS using CASA~\citep{CASA}, after providing the necessary metadata in the format expected by the \textsc{difx2fits} application.



First we use \sched\footnote{\sched: \url{http://www.aoc.nrao.edu/software/sched/}}, a program often used to schedule VLBI observations. Although scheduling is not necessary, the software produces the output files describing the details of the MeerKAT observation in a format that \difx\ can read. 
Hence, we first generate several files required to run \sched. This includes the station file with the location of the MeerKAT antennas that were used, a frequency file with the frequency setup, and the main KEY file with instructions for \sched. At this stage, we also generate the V2D file with information on the observing setup, Earth orientation and antenna clock offsets that will be used by \difx.
Once these files are created, we run \sched, which produces the VEX files that will be the input for \difx.  

We then run the \difx\ functions \texttt{vex2difx} and \texttt{calcif2} to produce a model of the geometric delays. We now have the delay model and \textit{uvw}-plane values required to assign to the \xgpu\ correlated visibilities. We next re-structure the \xgpu\ visibilites into a \difx\ format, including the metadata required such as the polarisation, band, and baseline. Finally, we use the \difx\ function \texttt{difx2fits} to convert the file into a FitsFile.

For every different number of baselines a new version of \xgpu\ needs to be compiled.
We would ideally always be observing with and saving data from all 64 MeerKAT dishes, however this is not always the case. To avoid compiling multiple versions of \xgpu,
we assume that we always have 64 dishes. To do this, we create fake antenna files prior to the \xgpu\ step that we can later flag.
For example, if we have only 60 dishes in an observations we create 4 copy antennas to pad to 64 dishes. Now that we have a \difx\ fits file,
we read this in to \casa\ using 
\texttt{importfitsidi}. 
We then use \casa\ \citep{the_casa_team_casa_2022} to flag the copied/fake antennas and the auto-correlations. Finally, we output the data as an MS that we can image.

\subsubsection{Producing images and transient localisation}

We now have one MS for each of the 64 frequency subbands. Since each MS technically has a different start time, instead of performing a joint deconvolution on all of the MSs together, we image each MS individually. We first perform a simple, dirty clean on each MS using \wsclean, and visually inspect the resulting images. This allows us to manually exclude channels dominated by RFI. This process will be automated in the future. We exclude those parts of the band that are dominated by RFI by excluding those MSs. We then produce a frequency and time average image by adding each dirty image together and dividing by the number of images. We compare this frequency and time averaged image to e.g. the ASKAP RACS-Mid \citep{duchesne_rapid_2023} of the same area of the sky to confirm that our image reflects reality. 

In order to detect an FRB we need to re-image each MS to produce images with shorter integration times. We expect the FRB to be close to the centre of the 300~ms due to the DM-slicing process, and therefore, we image in an odd number of time bins. We image in 11 time bins and proceed to average each time bin in frequency by adding the images in each bin together and dividing by the number of images. We now have one frequency averaged image per time bin. 

Each transient buffer dataset is 300~ms long, which means that the \textit{uv}-plane does not rotate significantly over the observation, and we do not expect the noise to change substantially over the dataset, even when taking into account the dispersion delay. This means that we can perform difference imaging to find the FRB. We do this by subtracting our time and frequency averaged image from each frequency averaged time bin image. We then visually inspect the resulting difference images to find the FRB. If we find the burst we confirm that it is the FRB by checking that it appears in the image corresponding to the FRB arrival time. 

Next we produce images with shorter integration times around the time bins where the FRB was seen, so that we can accurately select all the time bins where it was detected. We integrate these time bins to produce an ``on'' image, and then produce an ``off'' image with the same integration time where the FRB was not visible. We produce these images with more advanced cleaning parameters in \wsclean, which we also apply to the full integration time image.
The \wsclean~parameters we use for the stopping criteria are 100 iterations, or a threshold of 0.01 (arbitrary units). We apply a Cotton-Schwab cleaning with major iteration gain of 0.8, and auto-masking with $\sigma=3$. We apply a Briggs weighting with a robustness parameter of -0.3, and a weighting rank filter of 3. Finally, we use W-gridding on the data.


\subsection{Astrometry}

We corrected the absolute astrometry of the radio sources in the FoV of the detected FRBs using the method described in \citet{driessen_21_2022} and \citet{driessen2024}. We used the Python Blob Detector and Source Finder\footnote{\href{https://www.astron.nl/citt/pybdsf/}{https://www.astron.nl/citt/pybdsf/}} (\pybdsf) to determine the positions of sources in the full integration time, ``on'' and ``off'' images, which we used to determine and correct the accuracy of our absolute astrometry. 

For the astrometric corrections, where possible, we prioritised using reference catalogues that use Very Long Baseline Inferferometry (VLBI) to achieve milliarcsecond precision on the position, such as the Long Baseline Array (LBA) Calibrator Survey \citep[LCS1;][]{petrov_lba_2011}. Alternatively, the Australian Telescope Compact Array (ATCA) Parkes-MIT-NRAO (PMN) \citep[ATPMN;][]{mcconnell_atpmn_2012} has an astrometric accuracy of 0.4" in RA and DEC. 
However, these catalogues do not always have sufficient sources in the FoV of the images where the FRBs were localised.
The Rapid ASKAP Continuum Survey \citep[RACS;][]{hale_rapid_2021}, on the other hand, usually contains tens to hundreds of sources within the FoV, but the astrometric accuracy of the source positions has systematic offsets of $\sim1-2$ arcseconds due to the lack of sufficient radio sources with VLBI positions in the Southern Hemisphere to perform accurate astrometric corrections of the catalogue.

The Radio Fundamental Catalog (RFC\footnote{RFC: \url{http://astrogeo.org/rfc/}}), which provides positions with milliarcsecond accuracy, often contains more sources in the FoV of interest than LCS1 or ATPMN, but not enough to use on its own. 
When that was the case, we used RFC sources in a larger FoV than the image to correct the positions of the RACS sources, and finally used these corrected RACS positions to align the coordinates of the sources in the full integration MeerKAT images, using the \texttt{astroalign} module \citep{beroiz_astroalign_2020} in \textsc{Python}. We selected unresolved RACS sources with an uncertainty in both RA and Dec <0.5" and a total flux >20\,mJy. 

Once we obtained the transformation matrix for the full integration time image, we applied it to the ``on'' and ``off'' images and source positions to obtain the corrected FRB coordinates.
We computed the average separation between the corrected and reference sources after each alignment, and added them in quadrature to obtain the total astrometric error on the FRB position.
The details about the astrometric corrections we performed are given in Appendix~\ref{app:astrometry}.


\subsection{Offline beamforming}
Along with offline imaging, the channelized complex voltages saved to disk can be used to form beams at the best known location of the transient that is determined from the imaging and localisation. To do that, the corresponding gain solutions saved by the beamformer are used to phase up the interferometer to the phase centre of the observation. To form a phased beam at the location of the transient, one needs to multiply the gain/phase solutions by appropriate weights. In simple terms, this means adding an extra rotation phase to the existing vector of beamformed weights at the phase centre of the observation. To obtain these additional phase corrections, we use \textsc{MOSAIC}~\citep{chen2021} to compute the delay polynomials for each antenna i.e the expected delays that need to be added to each antenna to align the phase of the electric field from a given location in the sky. These are in-turn used to generate the beam weights as function of antenna and frequency. Since we extract the buffer data after compensating for the dispersion delay at each frequency channel, we generate the delay polynomials for each frequency separately based on the slightly different epoch of observation (corresponding to the dispersion delay at that frequency) before computing the weights. We note that these delays are similar to the delays computed during the imaging of these data and the differences are negligible. These weights are finally multiplied with the gain/phase solutions before they are applied to the channelized voltage data from the transient buffer. This process produces a phased up coherent beam at the location of the transient. Forming a coherent beam at the location of the transient has significant advantages: 1) the coherent beam contains all the antennas in the array unlike the core antennas typically used in the real-time search which increases the S/N of the detection  2) the coherent beams overlap at the 25$\%$ power point which means that FRBs that fall between two coherent beams get a significant boost (factor of $\sim$4) 3) the formed beam has the highest time-resolution possible with the correlator configuration and 4) there is polarization information available in the buffer data which can be used to study the polarization properties of the transient. The scripts used for offline beamforming are provided in an online repo~\footnote{\url{https://gitlab.com/kmrajwade/tbeamformer}}.

Before any scientific utilisation of polarization data can be done, it is important to take into account the effects of the primary beam on the polarization properties of the instrument. For a coherent beam that is pointing at a given location $x,y$~(where the origin is at the boresight of the primary beam) within the primary FoV, the measured electric field vector (for an elliptically polarized wave) for the electric field for each hand ($H$ and $V$) of polarization per antenna, per frequency channel,
\begin{equation}
\epsilon^{'}_{H,V}(x,y,i,\nu) = \mathcal{J}_{H,V}~\epsilon 0_{H,V}\, ,
\end{equation}
where the Jones Matrix,
\begin{equation}
\mathcal{J}_{H,V} = \begin{pmatrix}
j_{HH} &  j_{HV}\\
j_{VH} & j_{VV} 
\end{pmatrix}
\end{equation}
and the electric field vector,
\begin{equation}
\epsilon^{'}_{H,V}(x,y,i, \nu) = \begin{pmatrix}
E_{HH} \\
E_{VV} 
\end{pmatrix} .
\label{eq:jones}
\end{equation}

We assume here that for narrow channel widths, the electromagnetic wave can be considered to be monochromatic and thus, Jones algebra is applicable. Hence, in order to get the true measurement of the electric field at the position of the FRB, one has to correct for the primary beam Jones matrix. In order to obtain $\mathcal{J}_{H,V}$, we used the measurements from~\cite{devilliers2023} obtained from holography experiments with the MeerKAT telescope. Assuming that the Jones matrix for the primary beam does not change significantly with elevation, we use Eq.~\ref{eq:jones} to obtain the calibrated electric field for both hands of polarization. The resulting voltages are fully calibrated and can be directly used to measure the polarization of the detected FRBs.
We do note that this is not the most accurate method of calibrating the data as there is no measurement of the Jones matrix at the location of the FRB at the time of the FRB. We caution the reader that the correction may not entirely account for the leakage and we absorb these uncertainties with an additional $5\%$ uncertainty on the estimated polarization fraction.


\begin{figure*}
     \includegraphics[width=8.3cm]{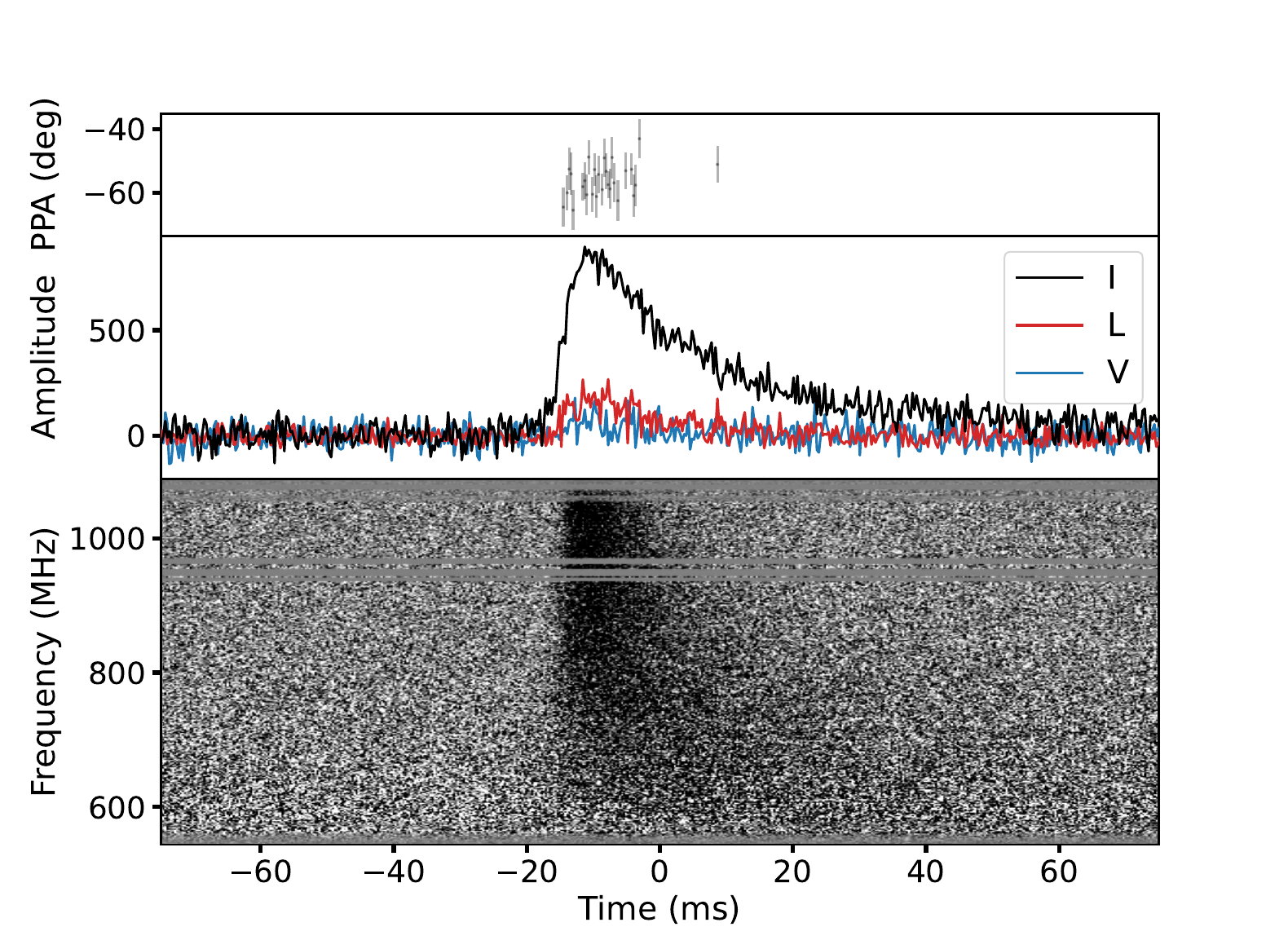}
     \includegraphics[width=8.3cm]{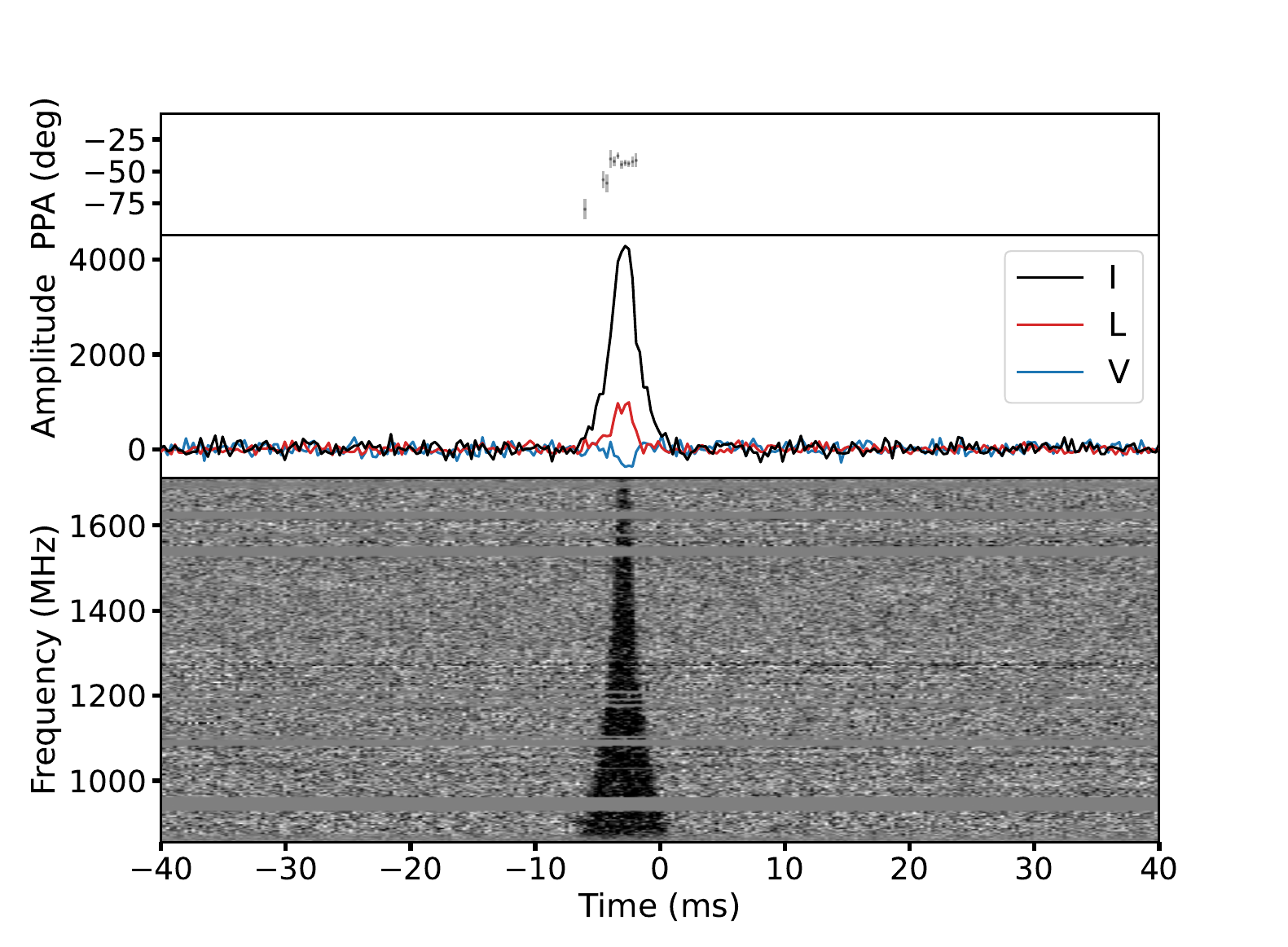}
    \caption{Calibrated polarization emission profiles for FRB~20220717A and FRB~20220905A created from the transient buffer data. The data for FRB~20220717A was dedispersed at the scattering-corrected DM while FRB~20220905A was dedispersed at a DM that accounts for the intra-channel DM smearing at the bottom of the band. The top panels show the absolute polarization position angle and the bottom panels show the total intensity (black), linear polarization (red) and circular polarization (blue).}
    \label{fig:example_figure}
\end{figure*}


\section{Results}

\begin{table*}
    \centering
    \begin{tabular}{lccc}
    \hline
    FRB parameter & Unit & FRB~20220717A & FRB~20220905A \\
    \hline
    MJD &  & 59777.8221507637 & 59827.7480359790 \\
    UTC &  & 2022-07-17T19:43:53.826 & 2022-09-05T17:57:10.309 \\
    RA (J2000) & (hms) & 19:33:13.0$\pm$0.9" & 16:54:19.8$\pm$0.7" \\
    Dec (J2000) & (dms) & -19:17:15.8$\pm$0.9" & -20:04:16.9$\pm$0.7" \\
    $l$ & (deg) & 19.83515767 & 0.78476176 \\
    $b$ & (deg) & -17.63203224 & 14.61426288 \\
    Detection frequency & (MHz) & 816 & 1284 \\
    S/N-maximising DM & (\dmunits) & $637.34\pm3.52$ & $800.61\pm0.60$\\
    Scattering-corrected DM & (\dmunits) & $634.69\pm0.10$ & -- \\
    Detection S/N && 15.3  & 14.4 \\
    Beamformed S/N && 101.1 & 141.9 \\
    $\tau_s$ \@ 1~GHz & (ms) & $8.2\pm0.3$ & -- \\
    Scattering index & & $-3.7\pm0.2$ & -- \\
    W$_{\text{50p}}^a$ & (ms) & $8.4\pm0.3$ & $1.1\pm 0.1$ \\
    W$_{\text{10p}}^a$ & (ms) & $20.1\pm0.6$ & -- \\
    W$_{\text{eq}}^a$ & (ms) & $10.0\pm0.3$ & -- \\
    RM &(rad~m$^{-2}$) & $385.7\pm0.4$ & $-83.1\pm1.9$ \\
    \hline
    $S_{\text{peak}}$ & (Jy) & $0.34\pm0.03$ & $6.40\pm0.04$ \\
    $F$ & (Jy ms) & $6.83\pm$0.03 & $7.0\pm$0.6 \\
    \dmne & (\dmunits) & 118 & 154 \\
    \dmymw & (\dmunits) & 83 & 104 \\
    \dmhalo & (\dmunits) & 86 & 115 \\
    \hline
    \end{tabular}
    \caption{Various observed and measured properties of FRB~20220717A and FRB~20220905A.\\
    $^a$ Measured at 1020.3~MHz.}
    \label{tab:frb_prop}
\end{table*}

\begin{figure*}
    \centering
    \includegraphics[width=0.49\textwidth]{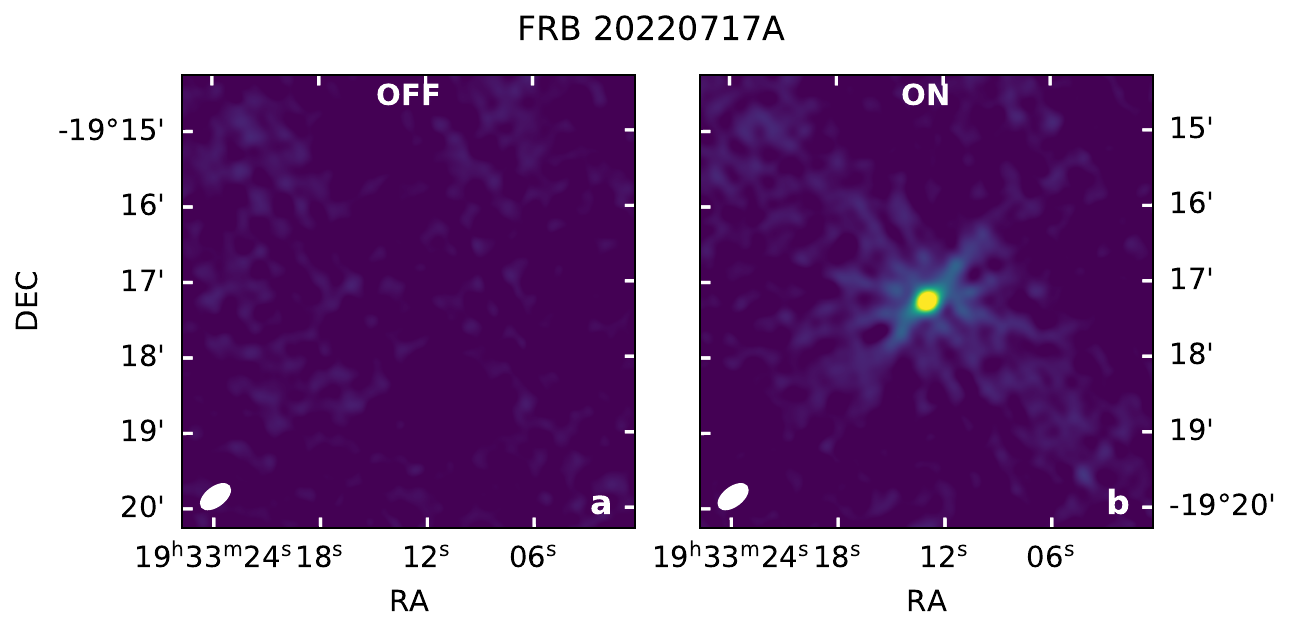}
    \includegraphics[width=0.49\textwidth]{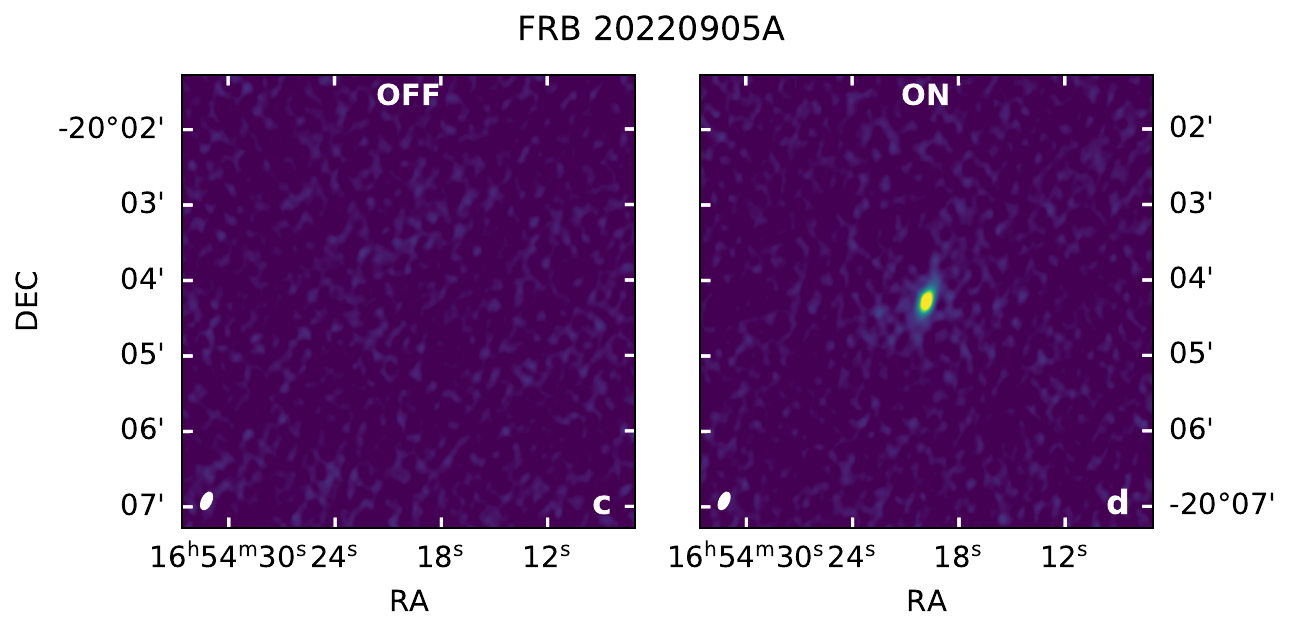}
    \caption{MeerKAT images of the localisation of FRB~20220717A and FRB~20220905A. Panel a shows a 43~ms integration of the region before the FRB detection (OFF), while panel b shows a 43~ms integration where the FRB was detected and localised (ON). The synthesised beam is shown on the lower left corner of each image. Panels c and d show similar images for FRB~20220905A for 7.7~ms integration.}
    \label{fig:frb20220717a_on/off}
\end{figure*}

\subsection{FRB~20220717A}
FRB~20220717A was discovered during commensal observations with the MeerTime project~\citep{bailes2020} at a DM of 637~pc~cm$^{-3}$. It was discovered at the UHF band (816~MHz) and shows clear evidence of scattering. The burst is broadband across the entire 544~MHz of bandwidth with no visible structure seen at smaller timescales. The burst shows a low linear polarization fraction (30$\pm$2$\%$) that maximizes at a rotation measure of 385.7$\pm$0.4~rad~m$^{-2}$. From the calibrated transient buffer data, we were able to localize the FRB to RA (J2000): +19:33:13.0$\pm$0.9" and DEC (J2000): $-$19:17:15.8$\pm$0.9" after performing an astrometric correction using the sources detailed in Table~\ref{tab:frb20220717a_astrom}. The errors on the position were obtained from summing in quadrature the \pybdsf error of the source position (0.4" RA, 0.4" Dec) and the error from the astrometric correction (0.9").

\subsection{FRB~20220905A}
FRB~20220905A was discovered during a MeerTime~\citep{bailes2020} observation at UTC 17:01:04. The FRB was detected at L-Band (1284~MHz) in the incoherent beam which triggered the storage of complex voltage data in the transient buffer. The FRB was detected at a DM of 800.6~pc~cm$^{-3}$ and shows no evidence of scattering or any emission at shorter timescales. Similar to FRB~20220717A, the FRB shows a low degree of linear polarization at a rotation measure of $-$83.81$\pm$1.9~rad~m$^{-2}$. The FRB was localized to RA (J2000): 16:54:19.8$\pm$0.7" and DEC (J2000): $-$20:04:16.9$\pm$0.7" which led to the immediate optical follow-up and identifying the the host galaxy as shown below.
The coordinates were obtained after performing an astrometric correction with the sources listed in Table~\ref{tab:frb_prop}. The errors on the position were obtained from summing in quadrature the \pybdsf error of the source position (0.09" in RA, 0.2" in Dec) and the error from the astrometric correction (0.7"). The astrometric corrections are detailed in Appendix~\ref{app:astrometry}.


\subsection{Optical Observations}
\label{sec:optical_obs}

\begin{figure*}
    \centering
    \includegraphics[width=\textwidth]{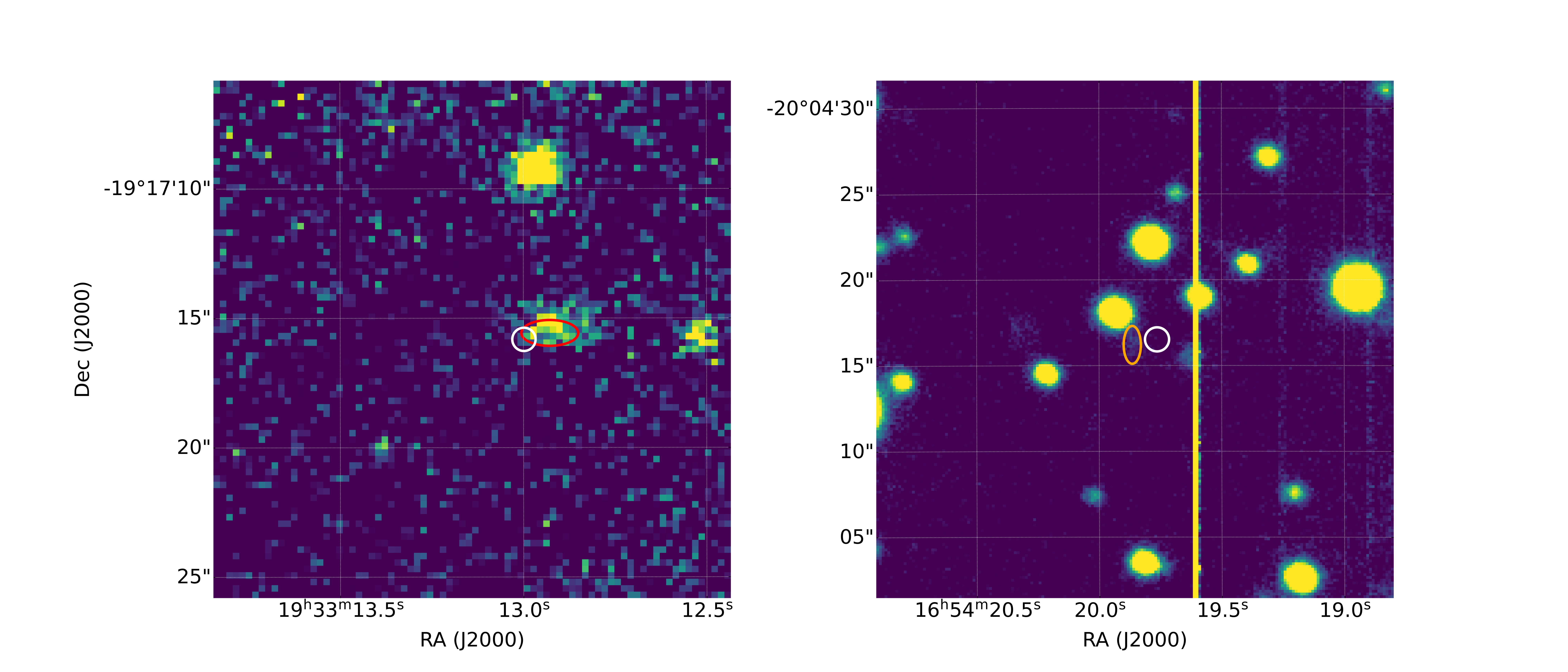}
    \caption{\textbf{Left:}Archival PanSTARRS DR1 image of the field surrounding the FRB20220717A localization (see PATH results in table \ref{table:path_2}). The best known 1$\sigma$ position of the FRB is shown by the white ellipse and the red ellipse shows the host galaxy. \textbf{Right:} GMOS image showing FRB20220905A localisation field crowded with stars (PATH results in table \ref{table:path_1}), white ellipse is 1 $\sigma$ localisation region, orange ellipse is the tentative host for FRB20220905A.}
    \label{fig:PS1}
\end{figure*}

We obtained deep imaging of the field of FRB\,20220905A with the Gemini Multi-Object Spectrograph on the 8-m Gemini South Telescope \citep{Gemini} to identify all possible host candidates (Program GS-2022B-Q-123, PI Gordon). We obtained 20x120s in $r$-band on 11 October 2022 UTC and 25x100s in $z$-band on 12 October 2022 UTC. Both datasets were reduced using the \texttt{POTPyRI}\footnote{https://github.com/CIERA-Transients/POTPyRI} pipeline. Then, we utilized the Probabilistic Association of Transients to its Host (PATH) method to link the transient to a host galaxy, as outlined in \cite{PATH_2021ApJ...911...95A}. We used photutils to perform photometry and found ~15 candidates within 30 arcseconds of the FRB localization (see Figure \ref{fig:PS1}). The prior that the host is unseen was set to be $P(U)=0.05$, and the offset prior was set to 50\% of the half-light radius of the host. PATH output indicated that the host is unseen (see \ref{table:path_1}). The PATH unseen posterior $P(U|x) \sim 1$. We also conducted a manual inspection of the image, during which we noted a faint $\approx 3 \sigma$ source 
offset 0.9 arcseconds from the FRB-localisation and with an angular size of 1.1 arcseconds. 
If we include this source in the list of candidates, 
it is assigned a very high PATH posterior ($P(O|x) \approx 0.98$). However, this candidate is still a tentative source.

 The FRB~20220717A localization is close (0.6") to a galaxy seen in PanSTARRS DR1 archival data of the field (see Figure \ref{fig:PS1}). A PATH analysis on the image confirmed the source (PSO J293.3038-19.2876) as the host galaxy of FRB~20220717A with a high posterior probability ($P(O|x)\approx 0.97$). We obtained spectroscopy of the host of FRB\,20220717A on 28 October 2022 UTC with the Goodman High Throughput Spectrograph on the 4-m Southern Astrophysical Research Telescope (SOAR; \citealt{SOAR}) to determine its redshift, totalling $2 \times 1200$\,s of science exposure (Program SOAR2022-007B, PI Gordon). We used the M1 400 lines/mm grating covering a wavelength range of 3000--7050\,\AA\ in conjunction with the BlueCam and a 1.0~arcsecond slit. The position angle was oriented to align the host with a nearby object for ease of identification during reduction. The data were processed with \textsc{PypeIt} \citep{pypeit}, using a quicklook reduction to identify the host redshift.

We obtained a second spectrum of the host of FRB\,20220717A with Keck/DEIMOS on 27 October 2022 UTC by taking a single 900s exposure (Program U129, PI Prochaska). We used the ZD 600 lines/mm grating for a wavelength coverage of 4550--9450\,\AA\ with a 1.0~arcsecond slit. The data were reduced fully using the PypeIt reduction package \citep{pypeit} to produce a flux-calibrated 1D spectrum of the host galaxy. This spectrum shows substantial contamination from skylines, likely due to a manufacturing issue during production of the relevant slit mask, which we were unable to remove fully in the reduction process. Nonetheless, we perform all further analysis on this DEIMOS spectrum. These observations yield a spectroscopic redshift of $z=0.36295 \pm 0.00018$ for the FRB host galaxy. 


\begin{figure}
     \includegraphics[width=0.5\textwidth]{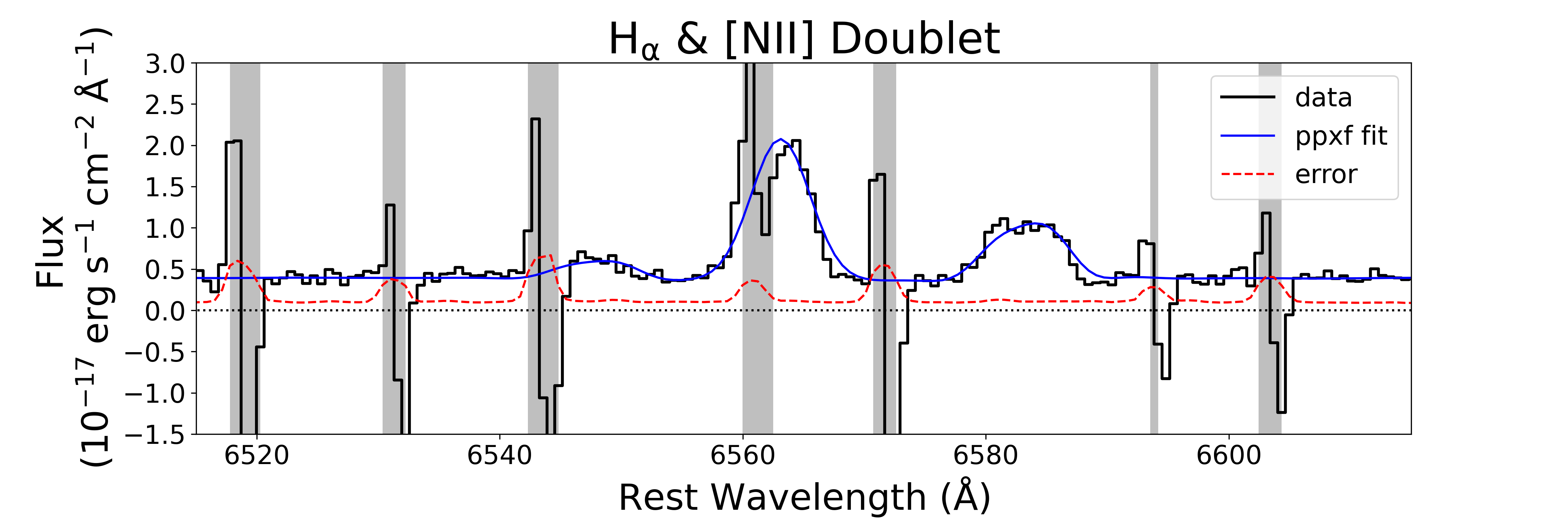}
    \caption{Keck/DEIMOS spectrum of the FRB20220717A host galaxy showing the H$_\alpha$ emission line at 6563\,Å and neighboring [NII] emission at 6548 Å and 6584 Å
    at a common redshift $z=0.3633$. 
    The black histogram shows the spectral data, while the observed error is shown in red. Blue shows the pPXF model fit to the data. Grey vertical regions indicate skylines that are masked in the spectral fitting process. 
    \label{fig:H_a-DM_L}}
\end{figure}

In order to confirm the redshift and to measure H$\alpha$ emission, the DEIMOS spectrum was fit using the \textsc{pPXF} spectral fitting package to fit both the spectral continuum and emission features \citep{Cappellari2023}. Due to the presence of poorly-subtracted skylines in the spectrum, we masked these regions out of the \textsc{pPXF} fit. Masking was applied to any region with a flux measurement error above 0.25$\times \mathrm{10^{-17}\;erg\;s^{-1}\;cm^{-2}}$\AA$^{-1}$, as well as any region with a flux measurement < 0.2$\times \mathrm{10^{-17}\;erg\;s^{-1}\;cm^{-2}}$\AA$^{-1}$ as this is indicative of over-subtraction. The resulting pPXF fit to the H$\alpha$ feature is shown in Figure \ref{fig:H_a-DM_L}. Integrating this fit yields an H$\alpha$ flux of 17.08 $\pm$ 3.7  $\times \mathrm{10^{-17}\;erg\;s^{-1}\;cm^{-2}}$, uncorrected for Galactic extinction.

Using the linear model to compute star formation rate (SFR) from H$\alpha$ emission given in \cite{Kennicutt1994}, this galaxy is observed to have an SFR of $0.65 \pm 0.14$ M$_\odot$\,yr$^{-1}$. Unfortunately this emission feature falls directly on an observed skyline, which was masked out of the flux integration measurements. Though we fit this feature using a Gaussian profile, the nearby [NII]$\lambda$6584 line shows a double-peaked profile indicating rotational broadening of the emission features. The limited data quality likely makes our measurement on H$\alpha$ an underestimation, and therefore our result for SFR computed therefrom should also be understood as a lower limit.

We estimate the host galaxy DM contribution using the H$\alpha$ emission measure (EM) as described in \cite{tendulkar2017}:
\begin{equation}
    \textrm{DM}_{\textrm{host}} = 387\ \textrm{pc\ cm}^{-3}\ L_{\textrm{kpc}}^{1/2} \left[ \frac{4f_{\textrm{f}}}{\zeta(1+\epsilon^2)} \right]^{1/2} \left( \frac{\textrm{EM}}{600\ \textrm{pc\ cm}^{-6}} \right)^{1/2},
\end{equation}
where $f_{\textrm{f}}$ is the volume filling factor of the ionized clouds, $\zeta \ge 1$ specifies cloud-to-cloud density variations, $\epsilon \le 1$ is the fractional variation within discrete clouds, and $L_{\textrm{kpc}}$ is the depth of the total ionized region in kpc. As in \cite{tendulkar2017}, we assume that $\zeta = 2$ (indicating 100\% variation between clouds) and that $\epsilon = 1$ (indicating that the electron density within clouds is fully modulated). We also assume that $f_{\textrm{f}}$ = 1. 

We compute EM from the observed H$\alpha$ surface brightness as described in \cite{Reynolds1977}. Because the PanSTARRS image of this host cannot be used to constrain its morphology, we cannot place good constraints on $L_{\textrm{kpc}}$. 
If we take $L_{\textrm{kpc}}$ to be 0.150, the expected value for a Milky Way-like spiral galaxy with the FRB in its midplane, we can thus estimate a $\textrm{DM}_{\textrm{host}}$ contribution of $\sim$ 100\,pc~cm$^{-3}$ \citep{Kalberla2009}. 


\section{Discussion}
\subsection{Benefits of complex voltage capture}
The ability to save complex voltage data from each antenna at the native time resolution allows MeerTRAP to overcome these limitations of post-detection analysis. Along with the ability to localize the FRBs by creating images from these data, we can also study FRBs at the finest possible time-resolution and also obtain polarization information. We also note that the ability to beamform the transient buffer data to the correct location of the FRB also enables one to increase the sensitivity of the telescope towards these FRBs significantly since the transient buffer data includes all the telescopes that were used in the observations as opposed to the limit of 40 dishes that is used in the real-time coherent searches for MeerTRAP. Furthermore, it also accounts for the reduction in S/N in the search due to offset of the FRB from the boresight of the coherent beam in which it was discovered. This is clearly shown by the difference in the estimated S/N of the bursts in the real-time search and the processed transient buffer data in Table~\ref{tab:frb_prop}. This ability enables one to reveal fainter features in the emission across the dynamic spectrum that may be washed out in the down-sampled data. The results presented in this paper reiterate the power of saving complex voltage data for FRBs.

\subsection{Complex environments around FRB progenitors}
 Both the FRBs presented here show flat polarization position angles (PAs). This is consistent with PAs observed for most one-off FRBs~\citep{pandhi2024}. It is important to note that PAs are also flattened due to scattering in the intervening medium based on observations of Galactic pulsars~\citep{li2003, karastergiou2009}. FRB~20220717A and FRB~20220905A show a very low degree of linear polarization (10$\pm2\%$ and 17.5$\pm1.5\%$) which is consistent with what has been recently seen for one-off FRBs~\citep{sherman2024, pandhi2024}. One of the possibilities of depolarization could be instrumental but any residual phase and gain differential between the two dipoles of the receiver will only increase the linear/circular polarization fraction hence depolarization is unlikely to be due to calibration inaccuracies. The calibration and leakage correction for circular polarisation measurements with MeerKAT is not yet well understood. As such, we could not reliably measure the circular polarization of these FRBs.

Recent studies of linear polarization of a large sample of FRBs~\citep{sherman2024, pandhi2024} have shown that one-off FRBs seem to have a large diversity in the degree of linear polarization. On the other hand, linear polarization fractions with values ranging between 90--100\%~\citep{mckinven2023} seems to be a distinct property of the repeating FRBs. These observations suggest a possible dichotomy in the nature of FRB progenitors, a key open question in the field. One argument for a small degree of linear polarization for some of the one-off FRBs could be a complex environment in the vicinity of the FRB source causing depolarization of radiation due to RM scattering~\citep{plavin2022}. Such environments could explain the large contribution by the host to the total DM in a number of apparently one-off FRBs recently discovered by ASKAP and MeerKAT~\citep{bhandari2023,caleb2019}. This might be true for FRB~20220717A with a potentially significant RM in the source frame but this conjecture is hard to reconcile with FRB~202209095A. To investigate the source of RM contribution, we compute the expected Galactic contribution to the RM along the line of sights of the two FRBs presented in this paper. To do this we use the Galactic RM maps created by~\cite{rm2022} to obtain the mean RM contribution by the Galaxy. The Galactic contribution along the line of sight to FRB~202209095A and FRB~20220717A is small (45 $\pm$ 17 and 0 $\pm$24~rad~m$^{-2}$), suggesting that the majority of the RM can be attributed to the host galaxy and any foreground, magnetized plasma.

\subsection{Origin of Scattering in FRB~20220717A}
\begin{figure}
        \centering
	\includegraphics[width=0.5\textwidth]{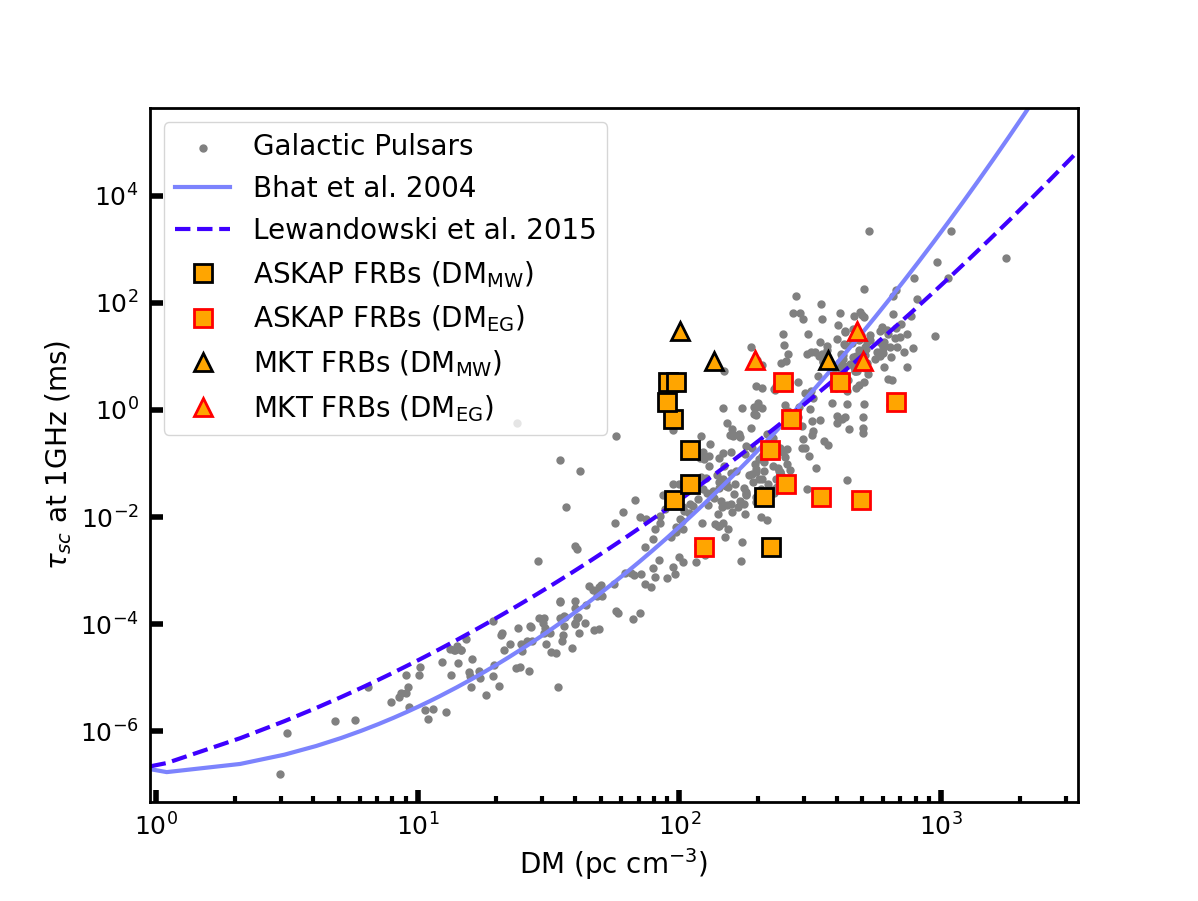}
        \caption{Dispersion measure versus scattering timescale at 1~GHz. The grey points show the measurement for
        Galactic Pulsars. The squares and triangles show scattering timescale as  a function of different DM 
        components for all ASKAP and MeerTRAP Localised FRBs. Here DM$_{\rm MW}$ refers to the DM contribution from the
        ISM of the Milky Way. Here we assume the MW halo contribution of 52.8~pc~cm$^{-3}$~\citep{cook2023}.}
\label{fig:scatcomp}
\end{figure}
\begin{figure}
    \centering
    \includegraphics[width=0.48\textwidth]{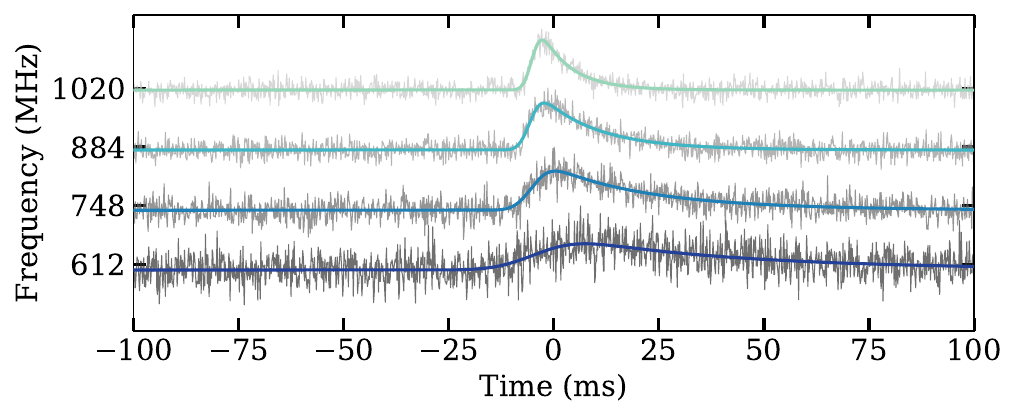}\\
    \includegraphics[width=0.48\textwidth]{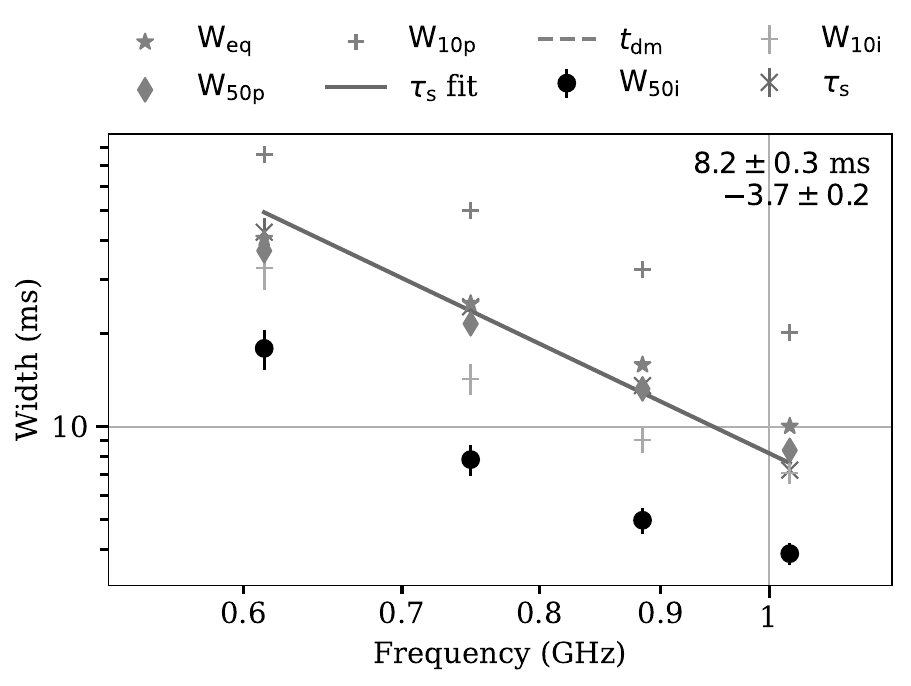}
    \caption{\textbf{Top Panel:} Scattered profile of FRB~20220717A shown in 4 subbands with the corresponding best-fit model. \textbf{Bottom Panel:} Estimate of scattering timescale and pulse width as a function of frequency along with the best-fit linear fit. The two number at the top right show the scattering timescale at 1~GHz and scaling index of scattering.}
    \label{fig:scattering}
\end{figure}

The burst from FRB~20220717A exhibits a strong scattering feature. In order to characterise it, we fit the profile with a combination of a Gaussian and a scattering function of the intervening medium. The scattering function can be approximated by an exponential quantified by the scattering timescale $\tau$. We use the \textsc{scatfit} software~\citep{jankowski2022ascl, jankowski2023} to fit the scattering function as a function of frequency. We split the data into 4 subbands such that there was enough signal in each to obtain a robust fit to the burst profile. Figure~\ref{fig:scattering} shows the results of our analysis. We obtain $\tau = 8.2 \pm 0.3$~ms at 1~GHz with the scattering timescale scaling with frequency as a power-law with an exponent, $\alpha = -3.7 \pm 0.2$. The analysis also optimises for the DM while fitting the scattering function so as to maximise the S/N which  gives us the best-fit DM of 634.69~pc~cm$^{-3}$.

The total DM of any FRB is made of different components such that,
\begin{equation}
\begin{split}
    \rm DM_{obs} = \dmism + \dmhalo + 
    DM_{EG}
    \\
    \rm DM_{EG} =  \dmcosmic  + \frac{DM_{host}}{1+z} \label{eq:dm}
\end{split}
\end{equation}
\noindent where \dmism\ is the contribution from the MW’s ISM and
\dmhalo\ is the contribution from the MW halo. $\mathrm{DM_{EG}}$ is the extragalactic DM contribution composed of \dmcosmic\ which is the contribution from the cosmic web (combined effects of the
intergalactic medium (IGM) and intervening galaxies), and
$\frac{\dmhost}{1+z}$ which is the redshifted contribution from the host galaxy's ISM including its halo and any gas in the immediate vicinity of the FRB source. An FRB with a known redshift allows us to estimate \dmcosmic\ and given that we can estimate \dmism\ and \dmhalo, we can then infer
an estimate for $\frac{\rm DM_{host}}{1+z}$. An increasing sample of accurately localized FRBs with identified host galaxies gives us an opportunity to assess the component of DM that contributes most significantly to the observed scattering in them. 

To do this, we collated all the well localized, scattered FRBs from ASKAP and MeerTRAP with measured redshifts~\citep{james2022, baptista2023, caleb2023, driessen_21_2022} and where the DM contribution from the host can be estimated based on the method presented in~\citep{james2022}. Then we looked for any correlations between the scattering timescale at 1~GHz and the DM contributions due to different components as shown in Figure~\ref{fig:scatcomp}. For majority of the FRBs, the expected scattering from the Milky Way for these FRBs is a lot smaller compared to the measured scattering timescale. It potentially hints at the fact that the measured scattering for FRBs cannot be explained by the ISM in our own Galaxy. Therefore, the scattering should be dominated by turbulence in the foreground galaxies and/or the host galaxy itself. For the published MeerKAT FRBs, it is evident that \dmhost\ can account for most of the scattering observed, further validating the claim made in~\cite{chawla2022}.


\subsection{Host Galaxy contribution to the DM of FRB 20220717A}

As discussed in the previous section, the dispersion of
FRBs makes them excellent probes for unraveling the 
structure of the cosmic web. This was initially shown 
by~\cite{macquart2020} who provided a relationship between
the expected DM$_{\rm EG}$ and the redshift of the FRB host 
(assuming a typical DM$_{\rm host}$ of 100 units). FRB~20220717A 
shows a host DM contribution that is consistent with these predictions. We compute the expected DM$_{\rm 
cosmic}$ for FRB~20220717A using the Macquart relation 
with the same assumptions as presented in~\cite{caleb2023}. Assuming a DM$_{\rm host}$ of 
100~pc~cm$^{-3}$ and a MW ISM and halo contribution of 83~pc~cm$^{-3}$  (from the \textsc{YMW16} model)
and 52~pc~cm$^{-3}$ using the model from~\cite{cook2023}, we obtain DM$_{\rm 
cosmic}$ of $\simeq$402~pc-cm$^{-3}$. This DM contribution from the IGM is within the scatter of the Macquart relation. It is worth noting that if we assume that the RM is mostly dominated by the host galaxy, the expected RM in the reference frame of the host, RM$_{\rm host frame}$ = RM(1+z)$^{2}\simeq$720~rad~m$^{-2}$. This is a large value of RM that is typically measured in the dense star-forming regions of a Galaxy~\citep{vaneck2021}. Furthermore, the high star-formation rate measured for the host galaxy can explain the turbulent and dense regions in the galaxy. These diagnostic measurements along with a large H$\alpha$ flux from the host galaxy spectrum suggests that FRB~20220717A may lie in a dense region of its host galaxy.

\section{Conclusions}
In summary, we present the discovery and the subsequent sub-arcsecond localisation of two FRBs with the MeerTRAP instrument. The transient buffer capture functionality has allowed us to localise and identify the host galaxies and study the polarization of these bursts. Both, FRB~20220717A and FRB~20220905A show a low degree of linear polarization with no conclusive evidence on the presence or absence of circular polarization due to calibration issues. This is consistent with what is observed for one-off FRBs and may hint at the fact that linear polarization fraction could be a distinguising property between the apparently repeating and non-repeating population of FRBs. It also suggests that there is a distibution in the polarization fraction in FRBs akin to single pulses seen from neutron stars and could be attributed to depolarization near the source. The host DM contribution for FRB~20220717A is estimated to be around 100~pc~cm$^{-3}$ which is consistent with the measured H$\alpha$ flux. The high star-formation rate of the host galaxy and the RM measurement suggests that the FRB may lie within a dense region of the galaxy. FRB~20220717A also exhibits scattering which can be mostly attributed to the host galaxy and the intervening medium, consistent with scattering seen in the FRB population. The transient buffer mode is fully operational on MeerTRAP with transient buffer data on more than 20 FRBs that are currently being investigated. This study again demonstrates the power of saving raw voltage data for accurately localising FRBs and further promotes the deployment of such systems on all real-time FRB detection systems around the world.

\section*{Acknowledgements}
The authors would like to thank the referee for their comments that significantly improved the manuscript. The MeerKAT telescope is operated by the South African Radio Astronomy Observatory, which is a facility of the National Research Foundation, an agency of the Department of Science and Innovation (DSI). The MeerTRAP collaboration would like to thank the MeerKAT Large Survey Project teams for allowing MeerTRAP to observe commensally. The MeerTRAP collaboration acknowledges funding from the European Research Council (ERC) under the European Union's Horizon 2020 research and innovation programme (grant agreement No 694745). The authors also acknowledge the usage of TRAPUM infrastructure funded and installed by the Max-Planck-Institut für Radioastronomie and the Max-Planck-Gesellschaft. KMR would like to thank Mattieu DeVilliers and Ludwig Schwarz for useful discussions regarding off-boresight polarization calibration for MeerKAT. The authors would like to thank the South African Radio Astronomy Observatory (SARAO) for immense support during the commissioning of the transient buffer mode. KMR acknowledges support from the Vici research programme ``ARGO'' with project number 639.043.815, financed by the Dutch Research Council (NWO). AM acknowledges support from the U.S. National Science Foundation through grant AST-2206492 and from the Nantucket Maria Mitchell Association. MC acknowledges support of an Australian Research Council Discovery Early Career Research Award (project number DE220100819) funded by the Australian Government. MB acknowledges support from the Bundesministerium für Bildung und Forschung(BMBF) D-MeerKAT award 05A17VH3 (Verbundprojekt D-MeerKAT).

\section*{Data Availability}

The processed data products from the transient buffer corresponding to FRB 20220905A and FRB 20220717A and the corresponding scripts will be made available to others upon reasonable request.



\bibliographystyle{mnras}
\bibliography{example} 




\appendix

\section{Astrometry sources} \label{app:astrometry}

Tables~\ref{tab:frb20220717a_astrom} and \ref{tab:frb20220905a_astrom} detail the sources that were used to perform the astrometric correction for FRB\,20220710A and FRB\,20220905A respectively. In both cases. no LCS1 or ATPMN sources were available, but several RFC sources laid in a 3.5\textdegree\ radius from the images phase centre. We thus used the RFC sources to align the positions of the matching RACS sources, and thus obtained the transformation to correct the RACS source position. Next we used the RACS sources matching the MeerTRAP sources obtained with \pybdsf\ to perform the final astrometric transformation. The resulting mean offsets between RFC and RACS, and RACS and MeerTRAP after each transformation, were added in quadrature to obtain the astrometric uncertainty. For FRB\,20220717A, this is $\Delta\theta=(0.20^2+0.83^2)^{1/2}=0.85"$, while for FRB\,20220905A, we get $\Delta\theta=(0.20^2+0.63^2)^{1/2}=0.66"$.
 
\begin{table}
    \centering
    \begin{tabular}{cccc}
    \hline
    RFC Source & RACS Source & Sep. before (") & Sep. after (") \\
    \hline
    J1924-1949 & J192441.4-194949 & 1.35 & 0.13 \\
    J1925-1813 & J192512.4-181303 & 1.44 & 0.36 \\
    J1928-2035 & J192809.1-203543 & 1.18 & 0.32 \\
    J1928-1707 & J192851.2-170758 & 1.71 & 0.20 \\
    J1930-2053 & J193010.3-205304 & 1.25 & 0.18 \\
    J1931-2025 & J193149.0-202537 & 1.11 & 0.19 \\
    J1935-1804 & J193509.3-180444 & 1.53 & 0.02 \\
    \hline
    & Mean & 1.36 & 0.20 \\
    \hline
    RACS Source & & Sep. before (") & Sep. after (") \\
    \hline
    J192139.2-175408 &  & 1.75 & 1.23 \\
    J192113.6-174846 &  & 1.60 & 0.98 \\
    J192109.9-170507 &  & 1.54 & 1.57 \\
    J192047.2-174602 &  & 1.26 & 0.60 \\
    J192045.1-164410 &  & 0.84 & 0.47 \\
    J192043.7-202838 &  & 1.91 & 0.83 \\
    J192043.6-185557 &  & 0.77 & 0.36 \\
    J192036.7-172940 &  & 2.28 & 0.87 \\
    J192036.5-202954 &  & 1.29 & 0.78 \\
    J192030.7-170746 &  & 0.80 & 0.26 \\
    J192032.7-191010 &  & 0.50 & 0.11 \\
    J192029.1-163509 &  & 1.79 & 1.01 \\
    J192017.5-174030 &  & 1.10 & 0.78 \\
    J192017.1-195115 &  & 1.03 & 0.63 \\
    J192008.7-203228 &  & 1.60 & 0.67 \\
    J192015.6-181904 &  & 3.29 & 0.86 \\
    J191941.1-180128 &  & 2.06 & 1.81 \\
    J191937.9-195826 &  & 1.80 & 0.86 \\
    J191919.7-205020 &  & 1.82 & 1.05 \\
    \hline
    & Mean & 1.53 & 0.83 \\
    \hline
    \end{tabular}
    \caption{Sources used for the astrometric correction of FRB\,20220717A. The first group were the RFC sources used to align RACS, while the second group are the corrected RACS sources used to align the MeerTRAP sources.}
    \label{tab:frb20220717a_astrom}
\end{table}

\begin{table}[h]
    \centering
    \begin{tabular}{cccc}
    \hline
    RFC Source & RACS Source & Sep. before (") & Sep. after (") \\
    \hline
    J1644-2156 & J164443.3-215608 & 1.15 & 0.03 \\
    J1647-1926 & J164753.7-192618 & 0.65 & 0.24 \\
    J1650-2010 & J165010.5-201012 & 0.93 & 0.24 \\
    J1656-2010 & J165655.1-201056 & 0.55 & 0.13 \\
    J1657-2004 & J165733.2-200434 & 0.88 & 0.25 \\
    J1701-2007 & J170135.4-200759 & 0.63 & 0.29 \\
    J1703-2110 & J170327.4-211049 & 0.70 & 0.22 \\
    \hline
    & Mean & 0.79 & 0.20 \\
    \hline
    RACS Source & & Sep. before (") & Sep. after (") \\
    \hline
    J165532.6-184546 &  & 2.45 & 0.33 \\
    J165204.9-212536 &  & 1.29 & 0.08 \\
    J165128.6-221213 &  & 0.99 & 0.49 \\
    J165118.8-231359 &  & 1.36 & 0.24 \\
    J165115.5-195629 &  & 1.30 & 0.12 \\
    J165059.1-230533 &  & 1.40 & 0.67 \\
    J165056.1-211911 &  & 0.28 & 0.58 \\
    J165054.4-232933 &  & 1.65 & 1.93 \\
    J165037.4-222326 &  & 1.26 & 0.57 \\
    J165033.9-201748 &  & 0.85 & 0.49 \\
    J164954.1-214558 &  & 0.07 & 0.26 \\
    J164953.0-220609 &  & 1.13 & 0.90 \\
    J164939.7-201149 &  & 1.01 & 0.85 \\
    J164910.4-183237 &  & 0.39 & 1.29 \\
    J164852.8-225423 &  & 2.11 & 2.19 \\
    J164846.4-214847 &  & 0.55 & 0.23 \\
    J164813.4-215206 &  & 1.74 & 0.83 \\
    J164753.7-192618 &  & 0.87 & 0.21 \\
    J164638.1-210942 &  & 1.64 & 0.11 \\
    J164528.9-195622 &  & 2.16 & 0.44 \\
    J164508.8-224833 &  & 1.40 & 0.89 \\
    J164438.8-184024 &  & 2.06 & 0.21 \\
    \hline
    & Mean & 1.27 & 0.63\\
    \hline
    \end{tabular}
    \caption{Sources used for the astrometric correction of FRB\,20220905A. The first group were the RFC sources used to align RACS, while the second group are the corrected RACS sources used to align the MeerTRAP sources.}
    \label{tab:frb20220905a_astrom}
\end{table}

\begin{table}
    \centering
    \begin{tabular}{cccccc}
    \hline
    RA & DEC & Ang-size & Mag & Sep & P(O|x) \\
    J2000 & J2000 & arc-sec & arc-sec & & \\
    \hline
    
    16:54:20.31 & -20:04:17.13 & 0.2 & 24.4 & 7.3 & $\ll$10$^{-6}$ \\
    16:54:19.81 & -20:04:03.52 & 0.3 & 21.1 & 12.7 & $\ll$10$^{-6}$ \\
    16:54:20.02 & -20:04:07.41 & 0.2 & 23.6 & 9.4 & $\ll$10$^{-6}$ \\
    16:54:19.20 & -20:04:07.62 & 0.2 & 22.7 & 12.0 & $\ll$10$^{-6}$ \\
    16:54:19.30 & -20:04:27.19 & 0.2 & 21.8 & 13.0 & $\ll$10$^{-6}$ \\
\hline
    \end{tabular}

\caption{PATH analysis results for FRB~20220905A showing the top 5 most probably host galaxy candidate. Here P(O|x) denotes the posterior probability of a galaxy being the host for the FRB. Most of the candidates have insignificant PATH posteriors. The value of 10$^{-6}$ roughly corresponds to 0.001$\%$ interval for a Gaussian probability density function.}
\label{table:path_1}
\end{table}

\begin{table}
\centering
    \begin{tabular}{cccccc}
    \hline
    RA & DEC & Ang-size & Mag & Sep & P(O|x) \\
    J2000 & J2000 & arc-sec & arc-sec & & \\
    \hline
    
    19:33:12.91 & -19:17:15.42 & 2.2 & 21.8 & 0.6 & $9.69 \times 10^{-1}$ \\
    19:33:13.12 & -19:17:09.35 & 2.2 & 20.9 & 5.6 & $2.28 \times 10^{-2}$ \\
    19:33:12.53 & -19:17:15.74 & 2.2 & 22.6 & 5.3 & $5.86 \times 10^{-3}$ \\
    19:33:13.17 & -19:17:30.54 & 2.2 & 24.9 & 16.2 & $7.37 \times 10^{-58}$ \\

\hline
    \end{tabular}
\caption{PATH analysis results for FRB~20220717A showing the most probable host galaxy candidates.}
\label{table:path_2}
\end{table}


\bsp	
\label{lastpage}
\end{document}